\documentclass[twocolumn,amsmath,amssymb,nofootinbib,floatfix]{revtex4}

\usepackage[pagewise]{lineno}
\usepackage{graphicx}
\usepackage{dcolumn}
\usepackage{bm}
\usepackage{hyperref}
\usepackage{comment}

\usepackage{color}
\usepackage{subfigure}

\newcommand{\Rmnum}[1]{\expandafter\@slowromancap\romannumeral #1@}

\def\be{\begin{equation}}
	\def\ee{\end{equation}}
\def\bea{\begin{eqnarray}}
	\def\eea{\end{eqnarray}}

\begin{document}
\title{Generation of axions and axion-like particles through mass parametric resonance induced by scalar perturbations in the early universe}

\author{Ruifeng Zheng}
\affiliation{Physics Department, College of Physics and Optoelectronic Engineering, Jinan University, Guangzhou 510632, China}
\author{Puxian Wei}
\affiliation{Physics Department, College of Physics and Optoelectronic Engineering, Jinan University, Guangzhou 510632, China}
\author{Qiaoli Yang}
\email[Corresponding author: ]{qiaoliyang@jnu.edu.cn}
\affiliation{Physics Department, College of Physics and Optoelectronic Engineering, Jinan University, Guangzhou 510632, China}

\begin{abstract}
Axions and axion-like particles can be generated in the early universe through mechanisms such as misalignment production, thermal processes, and the decay of topological defects. In this study, we show that scalar perturbations in the early universe can produce a significant amount of these particles primarily through mass parametric resonance effects. Scalar perturbations induce temperature fluctuations during the particle mass transition era, e.g., during the QCD phase transition. These temperature fluctuations modulate the particle mass, transferring energy into the field through parametric mass resonance, a nonlinear process. This mechanism exhibits substantially unstable regions that could lead to explosive particle production. Notably, it does not generate additional isocurvature perturbations.
\end{abstract}

\maketitle	
\section{INTRODUCTION}\label{(1)}
\par
The rapid development of astronomical and cosmological observations poses significant challenges to modern physics. One of the unsolved challenges is understanding the origin of dark matter. Observations indicate that a substantial fraction of the universe's matter is dark matter, accounting for approximately 27$\%$ of the total energy density \cite{Zwicky:1933gu, Arkani-Hamed:2008hhe, Bertone:2016nfn}. However, despite its significance, the composition of dark matter remains unknown. Various candidates spanning a wide range of energy scales have been proposed. For instance, primordial black holes (PBHs) have been studied \cite{Hawking:1971ei, Carr:1974nx, Cole:2023wyx, ZhengRuiFeng:2021zoz, Zhao:2023xnh, Qiu:2022klm}. In the energy range of approximately 10 GeV $\sim$ 10 TeV, weakly interacting massive particles (WIMPs) are among the leading candidates \cite{Bauer:2017qwy,BERTONE2005279}. Ultralight dark matter particles, such as axions and axion-like particles, represent wave-like dark matter candidates. Furthermore, light moduli \cite{Nakamura:2006uc, Endo:2006zj, Allahverdi:2010xz} are considered potential candidates \cite{ADMX:2018gho, Sikivie:2009qn, Duffy:2009ig, Arias:2012az, Chaudhuri:2014dla}.
\par
Peccei and Quinn proposed a solution to the Strong CP problem by introducing a global $U(1)$ symmetry, now known as $U(1)_{PQ}$ symmetry\cite{Peccei:1977hh, Peccei:1977ur}. This mechanism introduces a dynamic field, the axion, which absorbs the CP-violating angle $\overline{\theta }$ \cite{Weinberg:1977ma}. Currently, two benchmark axion models are widely studied \cite{Arvanitaki:2009fg, Kim:2008hd, Turner:1989vc, GrillidiCortona:2015jxo, Peccei:2006as}: the KSVZ model, proposed by Kim \cite{Kim:1979if}, Shifman, Vainshtein, and Zakharov \cite{Shifman:1979if}, and the DFSZ model, proposed by Dine, Fischler, Srednicki \cite{Dine:1981rt}, and Zhitnisky \cite{Zhitnitsky:1980tq}.

In addition to QCD axions, string theory predicts the existence of axion-like particles (ALPs) arising from extra-dimension compactification. These particles \cite{Svrcek:2006yi, Marsh:2015xka, Irastorza:2018dyq, Arias:2012az, Essig:2013lka, Jaeckel:2010ni} share properties similar to those of QCD axions. A major distinction is that ALP masses originate from the non-perturbative effects of string instantons, resulting in a broader parameter space compared with that of QCD axions. Both QCD axions and ALPs, along with other light particles, can be produced in the early universe through non-thermal mechanisms \cite{Preskill:1982cy, Abbott:1982af, Dine:1982ah, Kim:2008hd, Wantz:2009it, Graham:2015rva, Marsh:2015xka}. The production mechanism plays a crucial role in shaping the resulting dark matter particle ensemble. Consequently, exploring additional production mechanisms is essential for advancing our theoretical and experimental understanding of dark matter.

The misalignment mechanism is a fundamental process for generating light, wavelike dark matter particles in the early universe. This mechanism assumes that, in the early era, the particle field was effectively frozen owing to Hubble friction. Once the universe cooled below a critical temperature, the field's effective mass became comparable to values on the Hubble scale, triggering oscillations that converted potential energy into coherent particle excitations.

In scenarios where the field existed before inflation, it remained effectively massless and acted as a spectator during inflation. Consequently, the field underwent quantum fluctuations with a scale-invariant power spectrum. These fluctuations, proportional to $H_{inf}$, spatially modulated the initial field value. After dark matter particles are generated, energy transfer from the Standard Model sector to the dark matter field induces isocurvature perturbations. The absence of observed isocurvature perturbations in the cosmic microwave background places strong constraints on this scenario \cite{Hertzberg:2008wr, Hamann:2009yf, Wantz:2009it, Beltran:2006sq}.

In this study, we present a scenario in which scalar or pseudoscalar particles are generated through parametric resonance induced by scalar perturbations. Scalar perturbations, density fluctuations seeded by the inflaton, induce temperature fluctuations that modulate the axion mass during its mass transition era. The modulations can trigger parametric resonance, effectively injecting energy into the light fields. Resonance occurs when a system parameter, such as the mass, is periodically modulated near twice the natural frequency divided by an integer. Notably, this mechanism exhibits much larger instability regions than those arising from the linear resonance caused by external forces.

In addition, this process produces negligible isocurvature perturbations because it does not depend on the initial field value. However, residual isocurvature perturbations may still arise if the field existed before inflation and part of the particle ensemble was produced through the misalignment mechanism. Furthermore, while parametric resonance produces particles with non-zero momentum, these momenta are currently redshifted to negligible levels. Therefore, particles generated through this mechanism may be suitable dark matter candidates.

\section{Equation of Motion and field Potentials}\label{S2}
The scalar and pseudo-scalar fields $\phi(\vec{x},t)$ satisfy the following equation of motion in an expanding universe:
\begin{equation}\label{eq1}
 {D^{\mu } \partial _{\mu } \phi (\vec{x},t )-m^{2}(T(t)) \phi (\vec{x},t )=0~,}
 \end{equation}
where $D^\mu$ denotes the covariant derivative, and $m(t)$ represents the time-dependent particle mass, which evolves with the cosmological temperature $T(t)$. During the radiation-dominated era, the Hubble parameter is given by $H(t)=1/2t$. By considering only scalar perturbations and adopting the conformal Newtonian gauge for the space-time metric, one obtains
  \begin{equation}
 \label{eq2341}
 \begin{aligned}
 {-\partial _t^2\phi+\frac{1}{a^{2} } \partial_{j} ^2 \phi -3H\partial _t{\phi } - m^{2}(T(t))  \phi - f(\vec x,t,\phi)=0~.}
 \end{aligned}
 \end{equation}
The function $f(\vec x, t, \phi)$ is given by
\begin{equation}
\begin{aligned}
{f(\vec x,t,\phi)=}&{-2\Psi\partial _t^2\phi-\left(\partial _t\Psi-3\partial _t\Phi+6H\Psi\right)\partial _t\phi}\\
&{+\frac{2\Phi}{a^2}\partial _j^2\phi-\frac{1}{a^2}\partial _j(\Phi+\Psi)\partial _j\phi+\frac{dm^{2} }{dT}\delta T \phi}~,
\end{aligned}
\end{equation}
where $j = 1, 2, 3$ labels the spatial components. $a$ is the scale factor of the universe, and $\Psi$ and $\Phi$ are the metric perturbations. Assuming that the anisotropic stress in the primordial plasma is negligible, we have $\Psi=-\Phi$. During the radiation-dominated era and within the horizon $k\eta\gg1$, one finds \cite{Dodelson:2003ft, Mukhanov:2005sc}
\begin{equation}
\begin{aligned}
\Phi=&3\Phi_p(\vec{k})\left(\frac{\sin\left(k\eta/\sqrt3\right)-\left(k\eta/\sqrt3\right)\cos\left(k\eta/\sqrt3\right)}{\left(k\eta/\sqrt3\right)^3}\right)~~,
\end{aligned}
\end{equation}
and the temperature fluctuations are given by \cite{Dodelson:2003ft, Mukhanov:2005sc}
\begin{equation}
\begin{aligned}
\frac{\delta T}{T}=-\frac{3}{2}\Phi_{p}(\vec{k})\cos\left(\frac{k\eta}{\sqrt3}\right)~~~~~~,
\end{aligned}
\end{equation}
where $\Phi_{p}(\vec{k})$ denotes the primordial amplitude of $\Phi$ fluctuations. The conformal time is represented by $\eta$. The parametric modulations here can be roughly categorized into three components: the renormalization factor part, the friction part, and the mass part. Since the field mass increases during the phase transition, it is naturally the dominant contributor to the energy injection process.

\begin{figure}[!htbp]
	\centering
\includegraphics[scale=0.32]{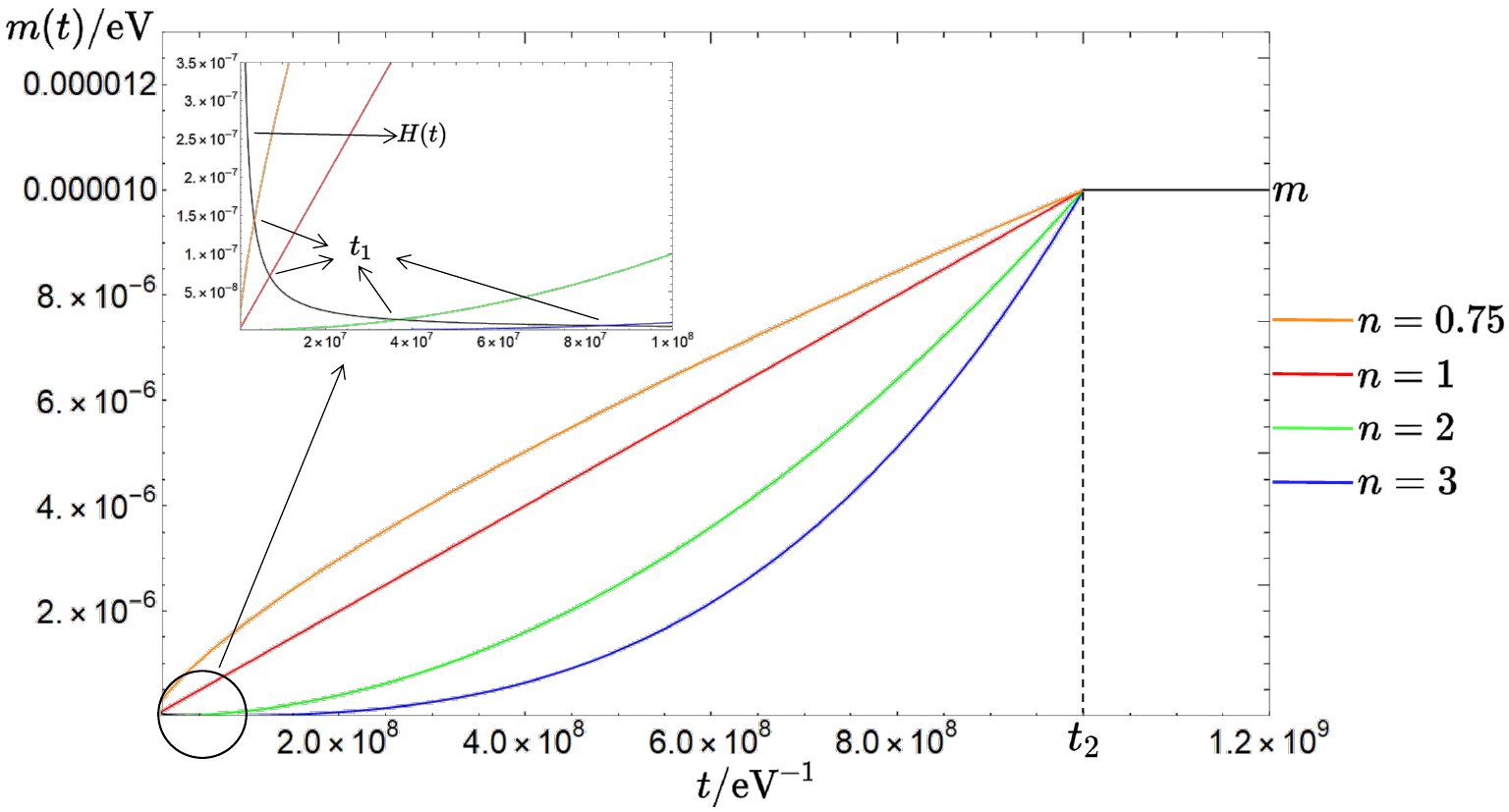}
	\caption{Mass as a function of time $t$.}
	\label{Fig1}
 \end{figure}

The general form of the light field potential as a function of temperature $T$ remains uncertain. In this study, we adopt the following phenomenological model:
 \begin{equation}
 \begin{aligned}
 \label{eqmass}
m(t)=\begin{cases}
 0~~~~~~~~~~~~~~~~t\ll t_1\\m\left(\frac{t}{t_2} \right)^{n}~~~~~t_1\le  t\le  t_2\\m~~~~~~~~~~~~~~~t> t_2~.
\end{cases}
 \end{aligned}
 \end{equation}
This model generalizes the case of \cite{Diez-Tejedor:2017ivd, Wantz:2009it, Bonati:2016imp} and also requires a smooth mass-temperature function (see Fig. \ref{Fig1}). We keep the exponent $n$ as a free parameter in the following equations to analyze its correlation with particle density.
Here, $t_1$ denotes the time at which the mass suppresses the hubble friction, and
\begin{equation}\label{axionmasstimess}
\begin{aligned}
t_1\simeq\left(\frac{t_2^{2n}}{m^2}\right)^{\frac{1}{2n+2}}~.
\end{aligned}
\end{equation}
The exponent $n$ determines the value of $t_1$, whereas $t_2$ represents the time at which the mass reaches its zero-temperature value $m$ (for QCD axions $t_2\sim 10^{-5}$s \cite{Husdal:2016haj})
\begin{equation}\label{t2m}
\begin{aligned}
t_2>t_1\simeq\left(\frac{t_2^{2n}}{m^2}\right)^{\frac{1}{2n+2}}\Rightarrow t_2>\frac{1}{m}~.
\end{aligned}
\end{equation}

\section{Field Dynamics}\label{S21}
Eq. (\ref{eq2341}) describes an oscillator with a variable frequency owing to temperature fluctuations. The periodic change in frequency leads to parametric resonance for modes with specific values of $k$. This phenomenon can be described by the Mathieu equation \cite{1968Theory, Kofman:1997yn}. Parametric resonance is more complex and powerful than linear (forced) resonance because the time-dependent parameters can induce exponential growth over a broader frequency range, whereas forced resonance yields linear growth within a narrow bandwidth. The equation of motion in $k$ space is
  \begin{equation}\label{foeq}
 \begin{aligned}
&\ddot{\phi}(\vec{k},t) + \frac{3}{2t} \dot{\phi}(\vec{k},t) + \left( k^{2} \frac{t_1}{t} + m^{2}(t) \right) \phi(\vec{k},t)\\
&-\frac{9 \Phi_{p}(\vec{k})}{2k^2tt_1}\ddot{\phi}(\vec k,t)  \cos \left( 2k \sqrt{\frac{ t_{1}t}{3}} t\right)
\\
&\quad+\Bigg[\frac{9\Phi_{p}(\vec{k})\sqrt{tt_1}}{\sqrt{3}kt^{2}t_1}\sin \left( 2k \sqrt{\frac{ t_{1}t}{3}} \right)\\
&\quad-\frac{27\Phi_{p}(\vec{k})}{4k^2t^2t_1}\cos \left( 2k \sqrt{\frac{ t_{1}t}{3}} \right)\Bigg]\dot{\phi}(\vec k ,t)
\\
&\quad+\Bigg[\frac{9\Phi_{p}(\vec{k})}{2t^2}-\frac{3\Phi_{p}(\vec{k})}{2}\frac{dm^2(t)}{dT}T(t)\Bigg]\\
&\quad\times\cos \left( 2k \sqrt{\frac{ t_{1}t}{3}} \right)\phi(\vec k, t)=0
~,
  \end{aligned}
 \end{equation}
where $a(t_1)=1$. The field evolution can be analyzed over two time periods: first, for $t_1<t<t_2$, during which the mass evolves with time, and second, for $t>t_2$, after which the mass no longer evolves and the last term in the equation vanishes. Equivalently, models with $k>k_2$ are not relevant for mass modulation resonances (see the definition of $k_2$ below). Parametric resonance for a given mode $\vec k$ occurs when the modulation frequency is close to twice the natural frequency divided by an integer $l=1,2,3...$.
\begin{equation}
{2\over l}\omega(k,t)={2\over l}\sqrt{k^2{t_1\over t}+m^2(t)}=m(t)+ k\sqrt{t_1\over3t}~.
\end{equation}
Since the mass no longer evolves after $t_2$, we define a critical wave vector $k_2$ corresponding to time $t_2$:
$k_2=m\sqrt{3t_2/t_1}$. In this context, we consider $l=2$, since $l=1$ has no solutions and $l$ with higher values are subdominant. The resonance time $t_R$ for a given $k$ mode is
\begin{equation}
\begin{aligned}
\label{111321}
t_R=\begin{cases}
 t_2\left(\frac{k}{k_{2} } \right)^{\frac{2}{2n+1} }~~~~\text{for}~k< k_2~\\
t_2\left(\frac{k}{k_{2} } \right)^{2}~~~~~~~~\text{for}~k> k_2~.
\end{cases}
\end{aligned}
\end{equation}
Eq. (\ref{foeq}) can be approximately reduced to the well-known Mathieu equation as follows:
\begin{equation}
\phi_{k}^{\prime\prime} + \left( A_{k} - 2q \cos 2z \right) \phi_{k} = 0~.
\end{equation}
During this stage, one may neglect terms such as the Hubble friction part, as such terms are significantly smaller than the mass term. Cosmic expansion effects can also be ignored. Defining $z=kt/\sqrt{3}$, Eq.(\ref{foeq}) becomes
\begin{equation}
\begin{aligned}
{\phi}''(\vec{k},z)+\bigg[3+\frac{3m^2(t)}{k^2}-2\times\frac{9\Phi_{p}(\vec{k})}{4k^2}\\\quad \times\frac{dm^2(t)}{dT}T(t)\cos(2z)\bigg]{\phi}(\vec{k},z)=0~,
\end{aligned}
\end{equation}
where the double prime denotes differentiation with respect to $z$. The parameters $A_k$ and $q$ in the Mathieu equation are given as follows:
\begin{equation}
\begin{aligned}
\label{l}
A_k&=3+\frac{3m^2(t)}{k^2}~;
\\
q&= \frac{dm^2(t)}{dT}\frac{9\Phi_{p}(\vec{k})T(t)}{4k^2}~.
\end{aligned}
\end{equation}
This corresponds to the $l=2$ resonance. In this case, the instability band is centered at $A_k\approx 2^2=4$. The driving frequency matches the natural frequency of the field. The width of the instability band satisfies $\Delta A_k\sim q^2$, corresponding to a fractional frequency bandwidth $\Delta\omega/\omega\sim \sqrt{1\pm q^2}$ (see Figs. 4, 5, and 6). While higher harmonics are possible, the contributions from smaller $l$ values dominate because they transfer more energy per cycle and more easily overcome damping.
\begin{figure}[!htbp]
\centering
\includegraphics[scale=0.30]{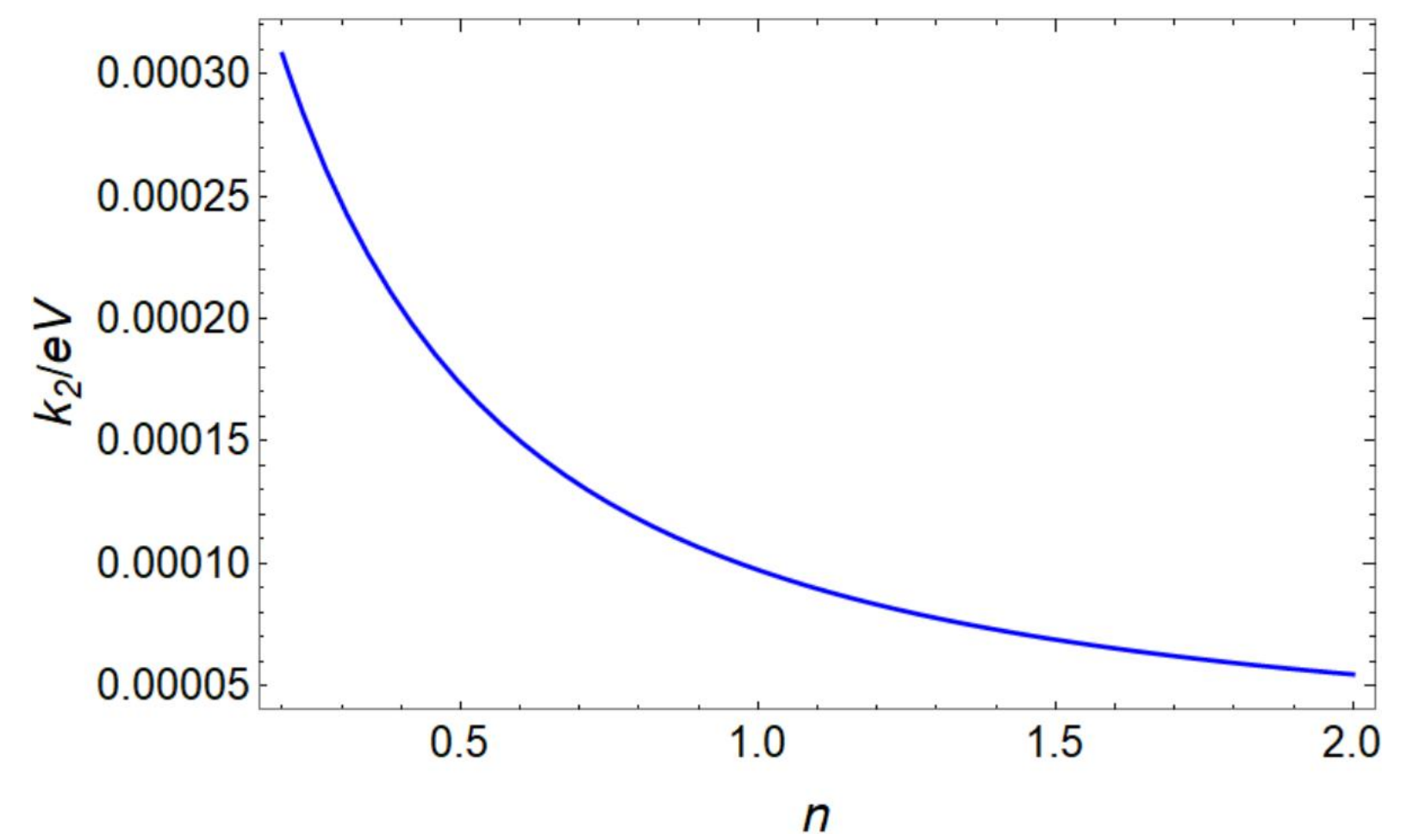}
\caption{Correlation between the parameter $n$ of the mass function and wave vector $k_2$ for $m=10^{-5}\ \mathrm{eV}$ and $t_{2}=10^{8}\ \mathrm{eV}^{-1}$.}
\label{Fig6}
\end{figure}
\begin{figure}[!htbp]
\centering
\includegraphics[scale=0.30]{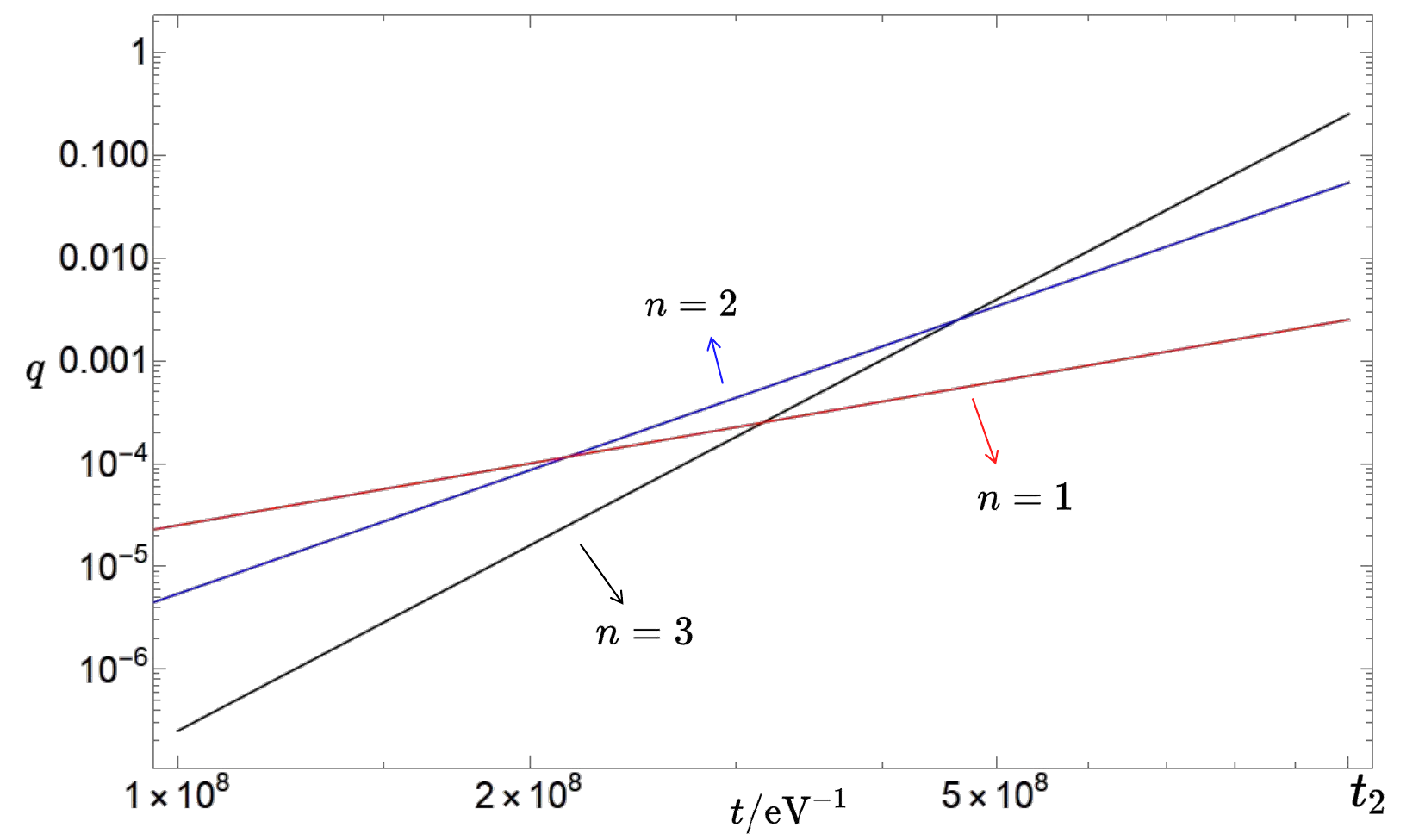}
\caption{Parameter $q$ in the Mathieu equation as a function of time $t$, with $l=2$, $m=10^{-5}\mathrm{eV}$, $t_2=10^{9}\mathrm{eV}^{-1}$, and $k=50/t_1$.}
\label{q}
\end{figure}
\begin{figure}[!htbp]
\centering
\includegraphics[scale=0.31]{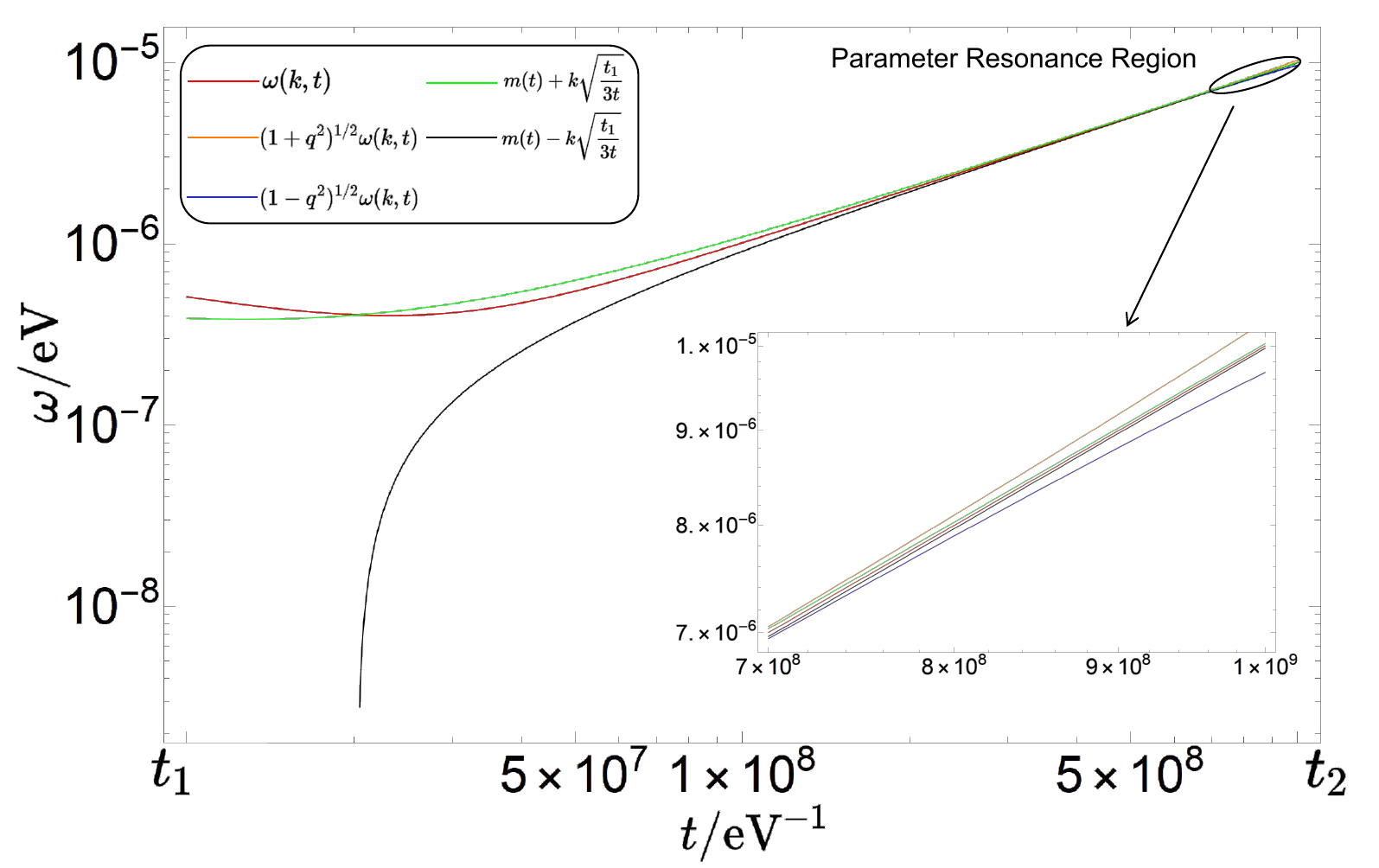}
\caption{Natural frequency $\omega$ and modulation frequencies $m\pm k\sqrt{t_1/3t}$ as functions of time $t$, with parameters $l=2$, $n=1$, $m=10^{-5}\mathrm{eV}$, $t_2=10^{9}\mathrm{eV}^{-1}$, and $k=5/t_1$.}
\label{n=1}
\end{figure}
\begin{figure}[!htbp]
\centering
\includegraphics[scale=0.39]{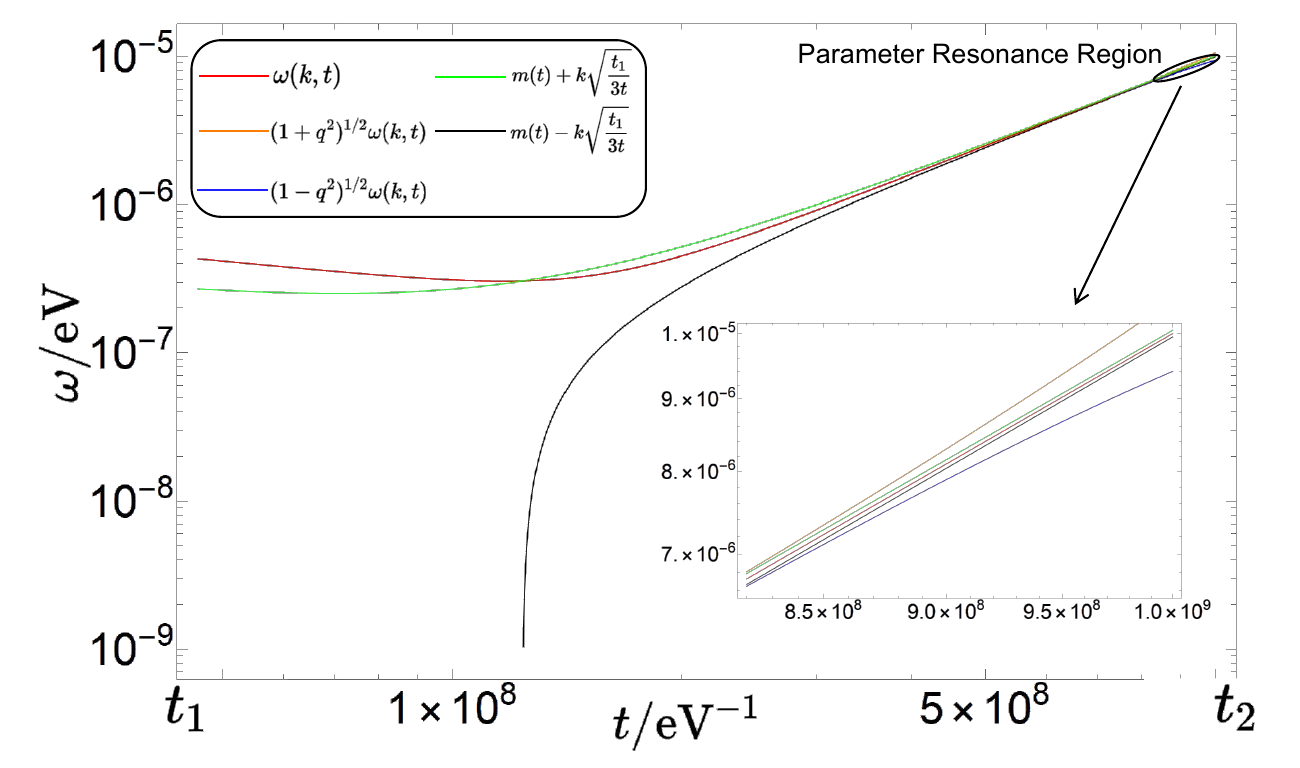}
\caption{Natural frequency $\omega$ and modulation frequencies $m\pm k\sqrt{t_1/3t}$ as functions of time $t$, with parameters $l=2$, $n=2$, $m=10^{-5}\mathrm{eV}$, $t_2=10^{9}\mathrm{eV}^{-1}$, and $k=20/t_1$.}
\label{n=2}
\end{figure}
\begin{figure}[!htbp]
\centering
\includegraphics[scale=0.34]{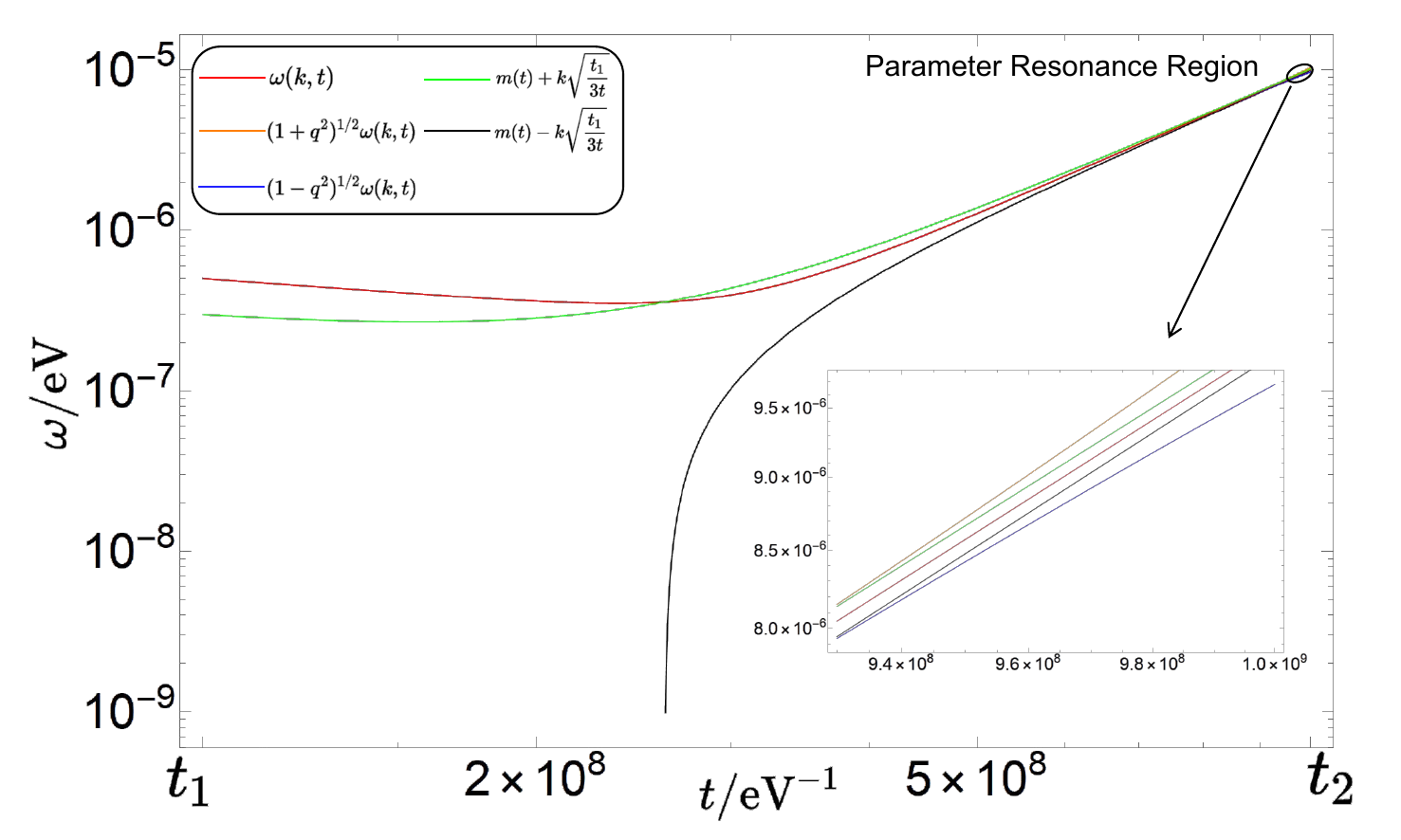}
\caption{Natural frequency $\omega$ and modulation frequencies $m\pm k\sqrt{t_1/3t}$ as functions of time $t$, with parameters $l=2$, $n=3$, $m=10^{-5}\mathrm{eV}$, $t_2=10^{9}\mathrm{eV}^{-1}$, and $k=50/t_1$.}
\label{n=3}
\end{figure}

Parametric resonance can only be solved analytically in specific cases, typically when the damping and driving terms are weak. Even in such cases, approximation methods are often required. Fortunately, in the "QCD axion-pump effect" scenario, where ground state QCD axions are excited to higher momentum states by primordial plasmons \cite{Sikivie:2021trt}, Eq.(\ref{foeq}) has been solved perturbatively. We adopt this solution to estimate the dark matter density produced. Note that, as $q$ approaches 1, the accuracy of the perturbative method diminishes however, the qualitative insights remain valid. More accurate predictions may require numerical methods in future studies.

One can decompose the field perturbatively as $\phi(\vec{x},t)=\phi^{0}(t)+\phi_{1}(\vec{x},t)$,
where $\phi^0$ corresponds to the initial field value (zeroth-order), and $\phi_1$ represents the component generated by parametric resonance. The zeroth-order equation is as follows:
\begin{equation}\label{foeq0}
\begin{aligned}
\ddot\phi^{0}(t)+3H(t)\dot\phi^{0}(t)+m^2(t)\phi^{0}(t)=0~,
 \end{aligned}
 \end{equation}
which can be solved using the WKB approximation. For $t\gg t_1$, the zeroth-order field is approximately
\begin{equation}
\begin{aligned}
\phi^{0}(t)=\frac{f_a}{\sqrt{m(t)t_1}}\left(\frac{t_1}{t}\right)^{\frac{3}{4}}\cos\left(\int_0^td{t}' m({t}' )+\delta\right)~,
\end{aligned}
\end{equation}
where $\delta$ is an initial phase. $\phi^0$ corresponds to a coherent particle ensemble with energy density \cite{Preskill:1982cy, Abbott:1982af, Dine:1982ah}
\begin{equation}
\begin{aligned}
\rho^0(t)=\frac{mf_a^2}{2t_1}\left(\frac{t_1}{t}\right)^{\frac{3}{2}}~.
\end{aligned}
\end{equation}
If the light field existed before inflation, this component would lead to isocurvature perturbations.

The first-order field $\phi_1$ can be evaluated perturbatively. For detailed calculations, see \cite{Sikivie:2021trt}, particularly the near-resonance region. The resulting energy density is
\begin{equation}
\begin{aligned}\label{eqrho}
\rho^{1}(t)=\frac{mAf_a^2}{4\pi^2}\left(\frac{t_1}{t}\right)^{\frac{3}{2}}\int^{k_2}_{1/t_1} dk\frac{|I(k)|^2}{k} ~,
\end{aligned}
\end{equation}
where $|I(k)|$ depends on the mass potential and the phase function. Here, $A\sim10^{-9}$ is the amplitude of the scalar perturbation power spectrum \cite{Planck:2018jri} related to $\left \langle \Phi_{p}\Phi_{p}^{*}\right \rangle$. With the mass model in Eq. (\ref{eqmass}),
\begin{equation}
|I|^2\simeq\frac{27n^2m^2t_2}{2t_1^2k^2}~.
\end{equation}
After redshift, the energy density of the particle ensemble becomes
\begin{equation}
\begin{aligned}
\rho^1(t)
&\simeq \frac{27Af_a^2n^2m^{\frac{6n+3}{2n+2}}t_2^{\frac{5n+2}{2n+2}}}{16\pi^2t^{\frac{3}{2}}}~.
\end{aligned}
\end{equation}
The energy density ratio compared with the zeroth-order part is
\begin{equation}\label{rhonr}
\begin{aligned}
\Omega_{1,0}&=\frac{\rho^1(t)}{\rho^{0}(t)}\simeq\frac{27An^2(mt_2)^{\frac{2n+1}{n+1}}}{8\pi ^2}~.
\end{aligned}
\end{equation}
\begin{figure}[!htbp]
\centering
\includegraphics[scale=0.57]{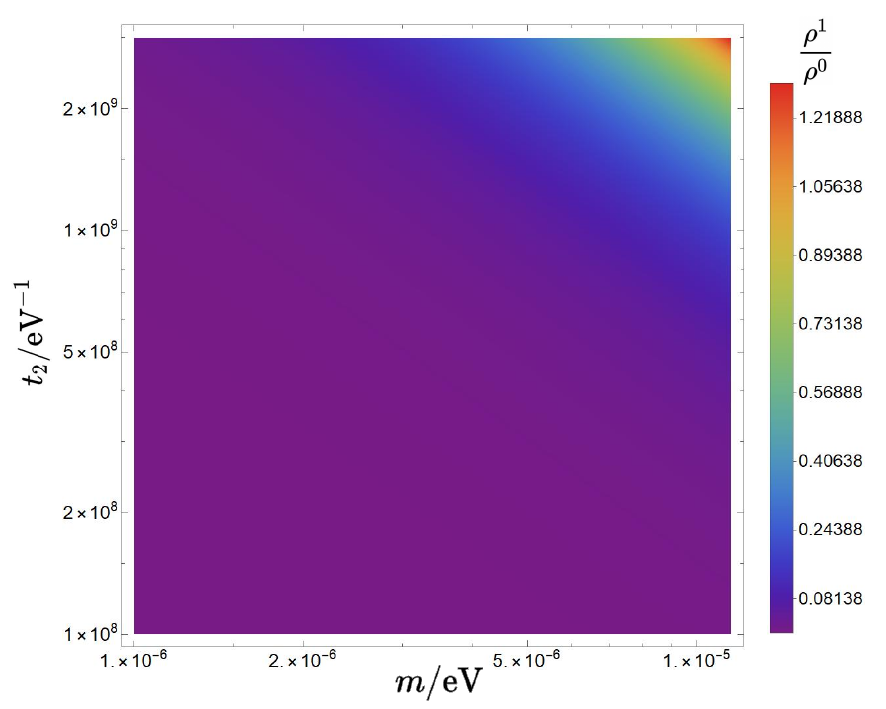}
\caption{Perturbative ratio between production from mass parametric resonance and misalignment production for $n=3$; see Eq.(\ref{rhonr}).}
\label{Fig11}
\end{figure}
Fig. (\ref{Fig11}) shows that the particle ensemble generated by parametric resonance could contribute a significant fraction, $\sim\mathcal{O}(1) $, to the total dark matter energy density. This contribution depends on the index $n$ and the timescale $t_2$ in the mass function. While this result holds when $q$ is small, it may underestimate the actual production when $q$ approaches 1 owing to stronger nonlinear effects.

\section{CONCLUSIONS}\label{S5}
The misalignment mechanism is one of the most important processes for the creation of light particles. If the corresponding light field existed before inflation, problematic topological defects, such as cosmic strings and domain walls, can be diluted. However, observations of isocurvature perturbations impose strong constraints on this scenario.

In this study, we show that scalar perturbations can generate light particles through parametric mass resonance. These particles are produced during the mass transition era, during which temperature fluctuations induce modulations of the particle mass. This is a nonlinear effect capable of efficiently transferring energy. The Mathieu equation reveals the presence of second-order $l=2$ instability regions. A perturbation estimate suggests that the amount of dark matter produced through this mechanism is comparable to that from the misalignment mechanism. The actual production may be greater owing to increasingly strong nonlinear effects later in the mass transition era.

Notably, this production mechanism does not induce additional isocurvature perturbations, even if the light field existed before inflation. Furthermore, since it depends solely on the modulation of particle mass during the mass transition era, the mechanism may also apply to particles beyond axions and ALPs.

\bibliographystyle{apsrev4-1}

\bibliography{ALP}

\end{document}